\documentclass[twocolumn,showpacs,superscriptaddress,preprintnumbers,amsmath,amssymb,longbibliography,prl]{revtex4-1}
\usepackage{graphicx}
\usepackage{dcolumn}
\usepackage{bm}
\usepackage{amsmath}
\usepackage{color}

\begin{document}

\title{Power-law scaling of early-stage forces during granular impact}

\author{Nasser Krizou}
\affiliation{Department of Physics, Naval Postgraduate School, Monterey, California 93943, USA}
\author{Abram H. Clark}
\affiliation{Department of Physics, Naval Postgraduate School, Monterey, California 93943, USA}

\begin{abstract}

We experimentally and computationally study the early-stage forces during intruder impacts with granular beds in the regime where the impact velocity approaches the granular force propagation speed. Experiments use 2D assemblies of photoelastic disks of varying stiffness, and complimentary discrete-element simulations are performed in 2D and 3D. The peak force during the initial stages of impact and the time at which it occurs depend only on the impact speed, the intruder diameter, the mass density of the grains, and the elastic modulus of the grains according to power-law scaling forms that are not consistent with Poncelet models, granular shock theory, or added-mass models. The insensitivity of our results to many system details suggest that they may also apply to impacts into similar materials like foams and emulsions.

\end{abstract}

\date{\today}


\maketitle	

High-speed impact by an intruder into a granular bed~\cite{vanderMeer2017} has broad relevance in many disciplines, including ballistics~\cite{Allen1957,Forrestal1992,Glossner2017,omidvar2019recent}, robotics~\cite{aguilar2016,aguilar2016review}, astrophysics~\cite{robbins2012new}, and earth science~\cite{Melosh1989}. The forces exerted by the grains on the intruder are often well described by Poncelet models, which are dominated by a velocity-squared drag force~\cite{durian_nat07,goldman_pre08,umbanhowar_pre10,bless2018}. However, the initial impact forces are consistently larger than expected from these models~\cite{picaciamarra_prl04,bless2012deceleration,clark_epl13,clark_pre14,tiwari2014drag,bester_pre17,Glossner2017,bless2018}. Very little is known about these early-stage forces, particularly in the high-speed regime where the impact speed, $v_0$, approaches a characteristic force propagation speed in the granular bed, $v_b$. In this Letter, we use experiments and simulations to study the early-stage forces in this regime. We find that the peak forces and associated times obey simple, power-law scaling forms that depend only on the impact speed, the intruder diameter, the mass density of the grains, and the elastic modulus of the grains. These scaling laws do not fit within the framework of any existing theory related to impact, including Poncelet models, granular shock theory, and added-mass models.

\begin{figure}
\raggedright \hspace{5mm} \hspace{80mm} \\ \centering 
\includegraphics[trim=0mm 0mm 0mm 0mm,clip,width=\columnwidth]{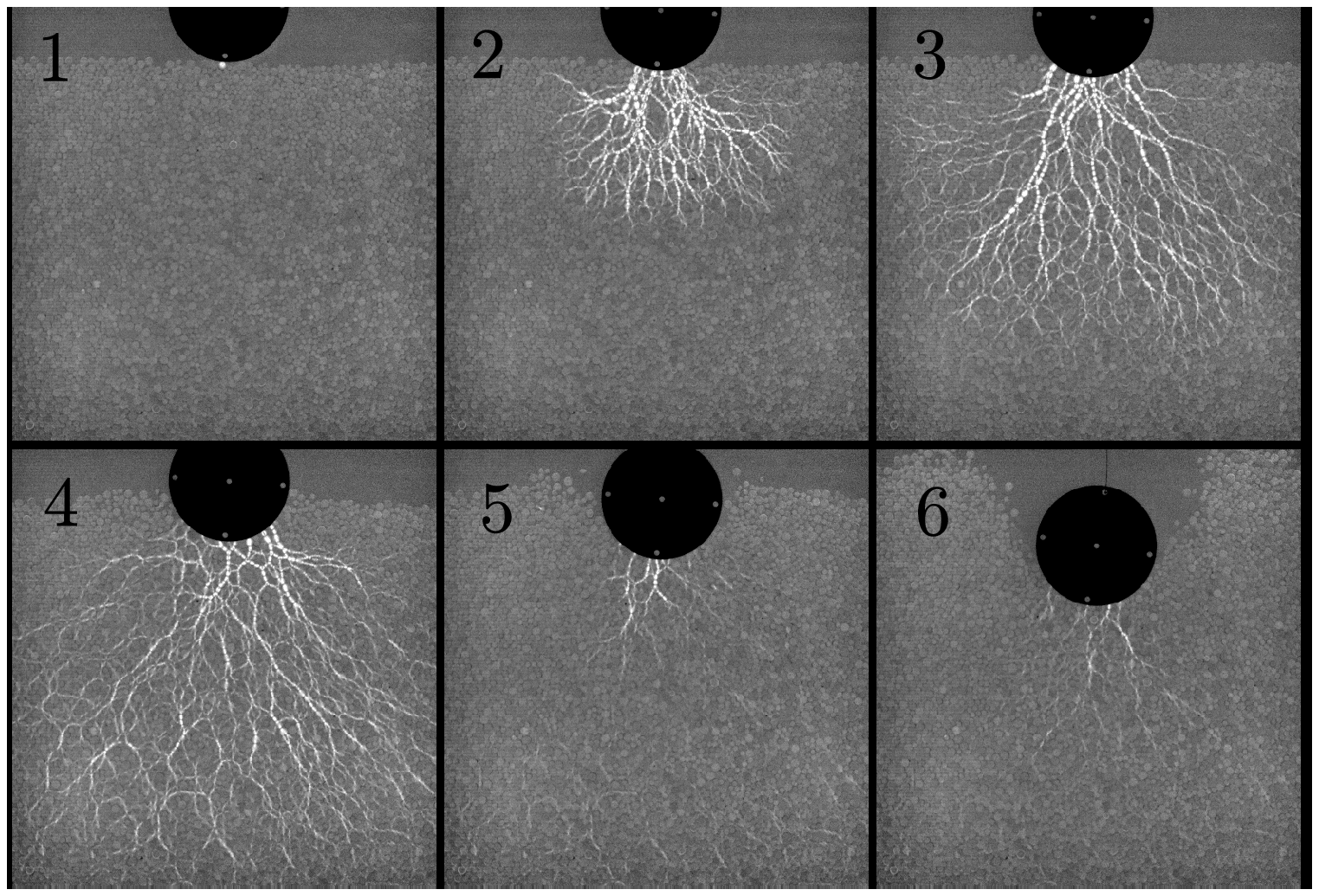}\\
\includegraphics[trim=0mm 0mm 0mm 0mm,clip,width=\columnwidth]{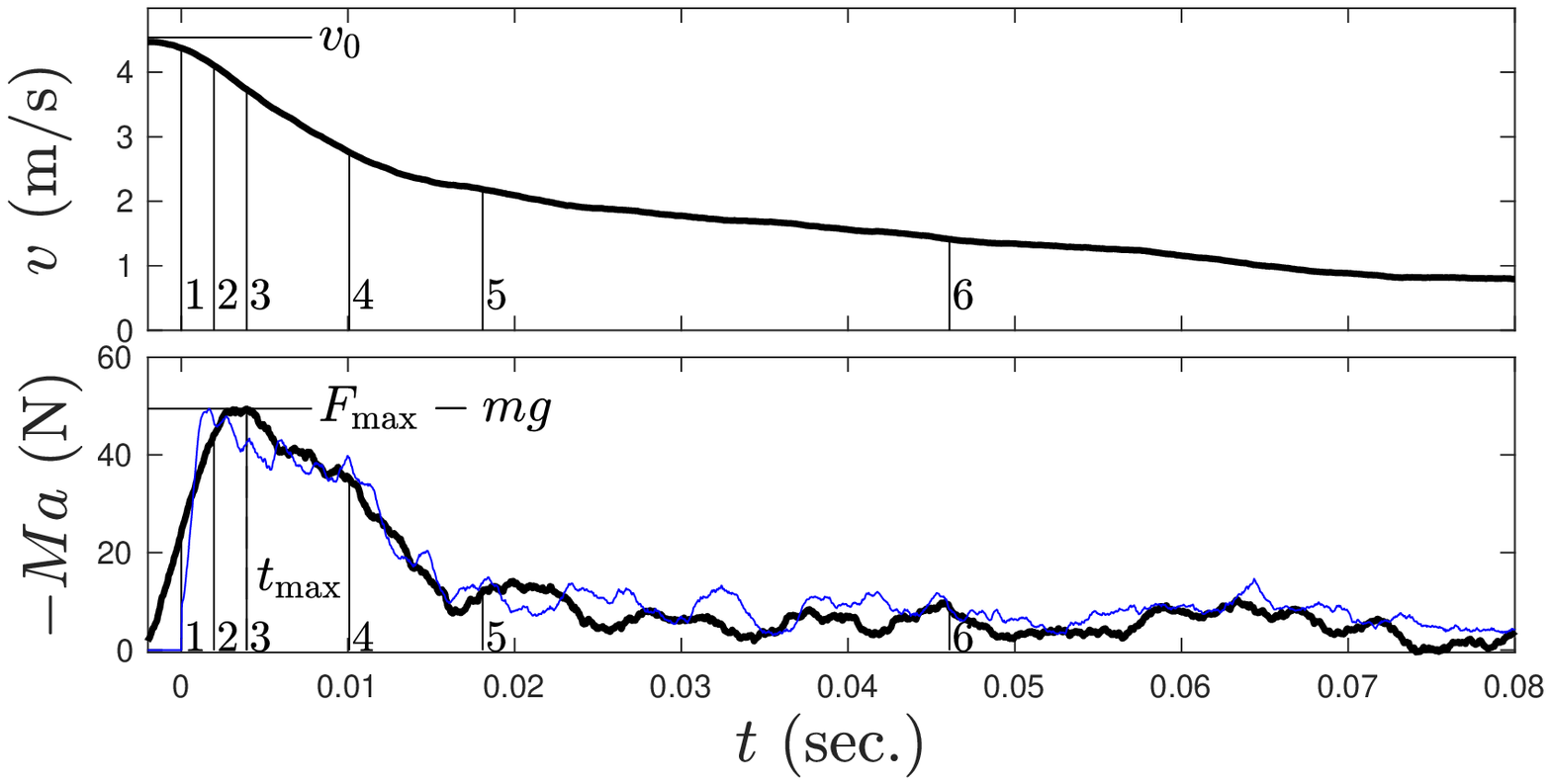}

\caption{Photoelastic image sequences are shown for an impact of the bronze intruder with $D=12.7$~cm at $v_0\approx 4.4$~m/s into medium particles (soft particles are shown in Supplemental Material), with corresponding trajectories shown below. 
Times marked 1-6 correspond to images 1-6 shown above. The thick black curve comes from tracking the intruder, and the thin, blue curve is a time series of the total, calibrated photoelastic response in a region beneath the intruder.}
\label{fig:soft-frames}
\end{figure}

Experiments involve circular intruders falling due to gravity and striking a collection of more than 10,000 photoelastic disks. These experiments have been used previously to study the microscopic origins of Poncelet drag~\cite{clark_prl12,clark_epl13,clark_pre14} as well as the speed and spatial structure of the propagating forces~\cite{clark_prl15}. Intruders are bronze disks with diameters $D=6.35$, 12.7, and 20.32 cm and masses of $M = 0.062$, 0.258 and 0.671 kg, respectively. We also cut one circular intruder out of aluminum with diameter $D=12.7$~cm with $M = 0.076$ kg. Photoelastic particles are made from three different materials of varying stiffness, as described in Ref.~\cite{clark_prl15}. For all particles, the force $f$ required to compress a particle by a distance $\delta$ is experimentally found to obey $f=E^\ast w d\left(\frac{\delta}{d}\right)^\alpha$, where $\alpha\approx 1.4$, $w = 3$~mm is the particle thickness, $d$ is the particle diameter, and $E^\ast$ is an effective Young's modulus. We find $E^\ast \approx$~3 MPa, 23 MPa, and 360 MPa for soft, medium, and hard particles, respectively. We use bidisperse mixtures: hard particles have $d=4.3$ and $6$~mm, and medium and soft particles have $d=6$ and $9$~mm. We find bulk densities $\rho_g \approx 1100$ kg/m$^3$ for all three types of particles. The velocity scale for propagating forces is set by $v_b = \sqrt{E^\ast / \rho_g}\approx 52$ m/s, 145 m/s, and 572 m/s for soft, medium, and hard particles, respectively~\cite{clark_prl15}. Initial impact speeds are $0.3 < v_0 < 6$ m/s, meaning that we experimentally access impact speeds in the range $10^{-3}<v_0/v_b<10^{-1}$. We record results with a Photron FASTCAM SA5 at frame rates of 10,000, 25,000, and 40,000 frames per second for soft, medium, and hard particles, respectively. Intruder trajectories are determined by tracking the position $z$ of the intruder ($z=0$ represents the impact point, and downward is positive $z$) at each image; sample images and trajectories are shown in Fig.~\ref{fig:soft-frames}. Since discrete differentiation of noisy data requires a low-pass filter, we cut off some high-frequency data in the intruder velocity $v=dz/dt$ and acceleration $a=d^2z/dt^2$, and we use a calibrated photoelastic signal as a secondary source of force information. 

Figure~\ref{fig:soft-frames} shows photoelastic images along with corresponding trajectory plots during impacts into medium particles (a video can be found in the Supplemental Material of Ref.\cite{clark_prl15}). We observe similar phenomenology for soft particles, as shown in Supplemental Material. The force exerted by the granular material onto the intruder shows a clear buildup to a maximum $F_{\rm max}$ on the intruder at time $t_{\rm max}$ (frame 3). After this, the shock wave continues to propagate down into the material, but the force on the intruder begins to relax. The force from tracking the intruder (thick black curve) and from the photoelastic signal (thin blue curve) show good agreement, confirming that we are not missing any high-frequency dynamics from the lowpass filtering of the trajectory. System boundaries do not affect our results: the boundary of the experiment is roughly 5 particle diameters below the bottom of the image, and the sidewalls are more than 20 particle diameters outside the edges of the image shown, so the shock front at $t_{\rm max}$ {has not reached the boundaries in any direction}. We also verify that boundaries do not affect our results using simulations by varying system size over a large range.

\begin{figure}
\raggedright (a) \\ \centering
\includegraphics[trim=0mm 0mm 0mm 0mm, clip, width=\columnwidth]{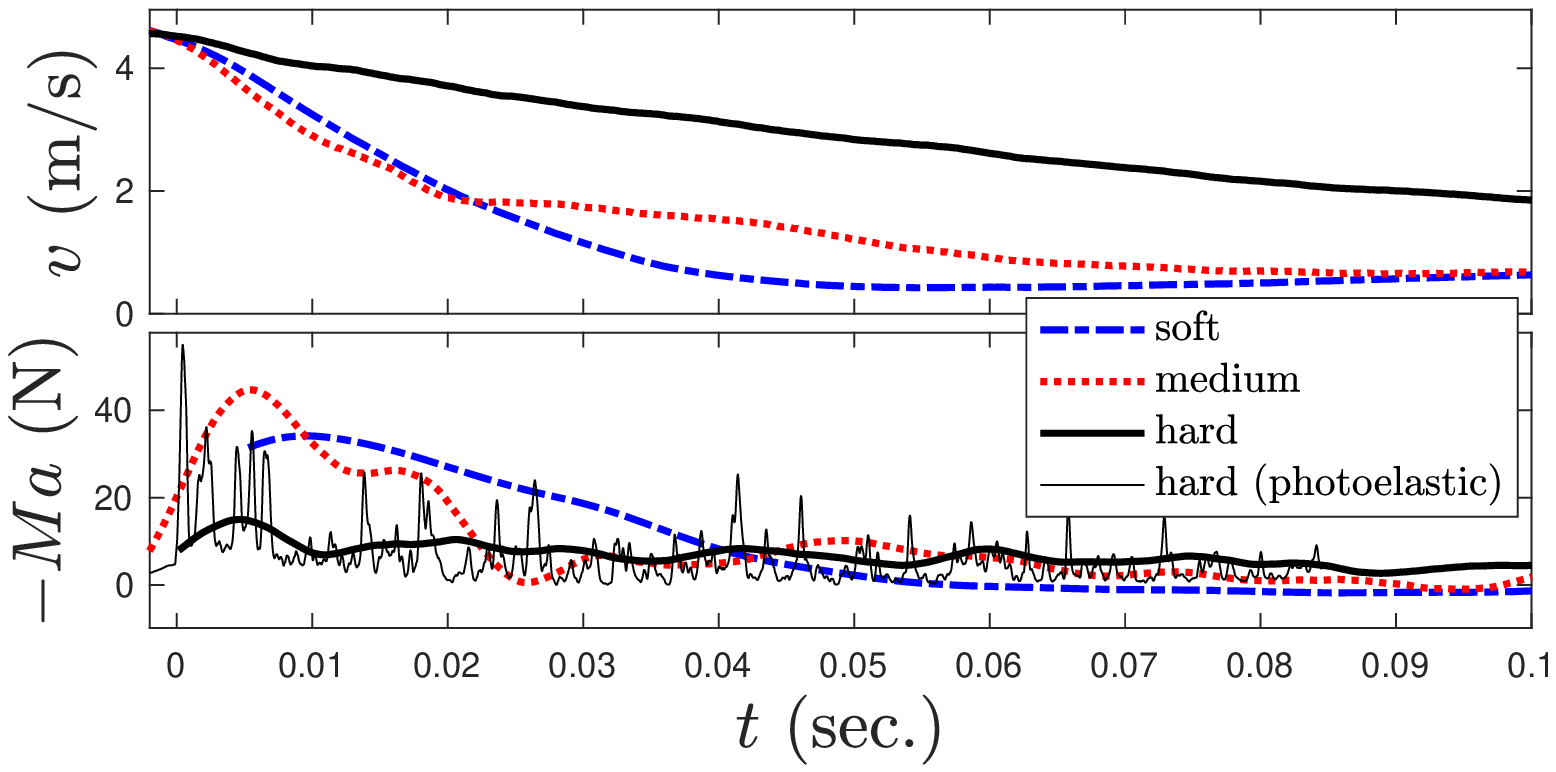} \\
\raggedright (b) \hspace{40mm} (c) \\ \centering
\includegraphics[trim=0mm 0mm 5mm 0mm, clip, width=0.49\columnwidth]{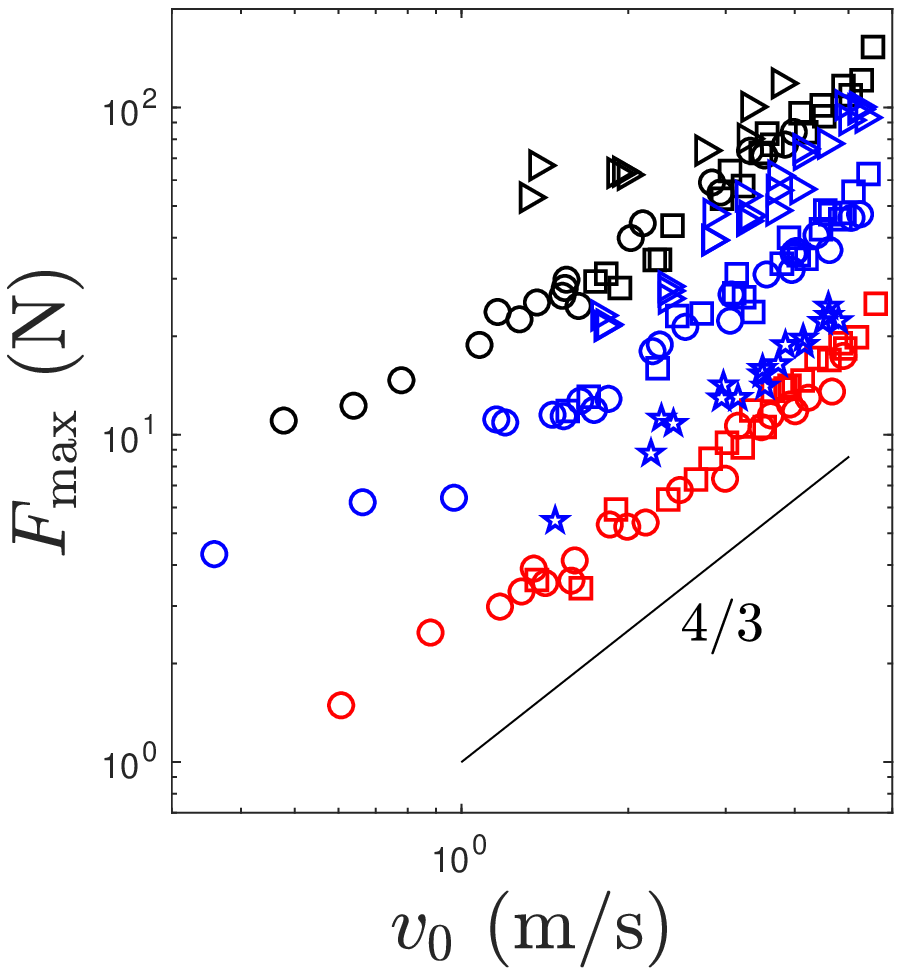}
\includegraphics[trim=0mm 0mm 5mm 0mm, clip, width=0.49\columnwidth]{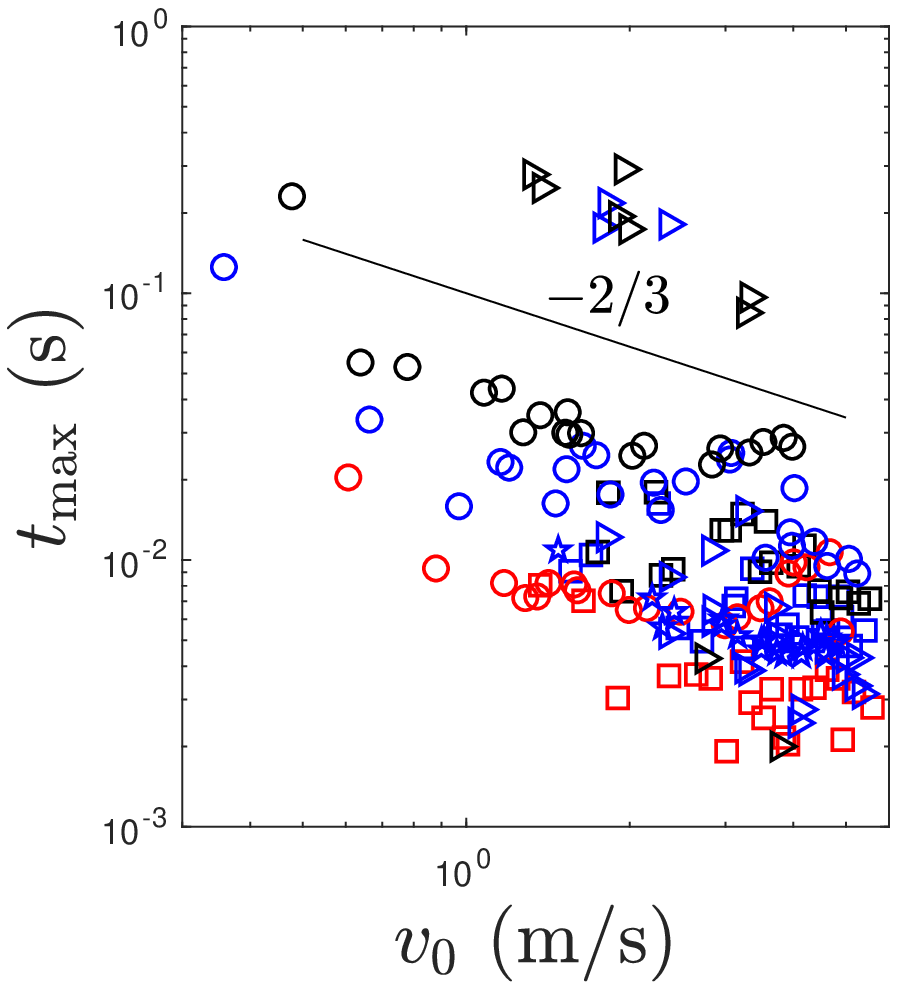}
\caption{Panel (a) shows sample trajectories for the bronze intruder with $D=12.7$~cm impacting into hard, medium, and soft particles. The photoelastic force is obtained by calibrating a time series of the photoelastic response beneath the intruder~\cite{clark_prl12}. (b) $F_{\rm max}$ and (c) $t_{\rm max}$ are then plotted as a function of $v_0$. Circles, squares, and triangles represent soft, medium, and hard particles (respectively) with bronze intruders. Stars represent impacts of aluminum intruders into soft particles. Red, blue, and black represent intruder diameters $D=6.35$, 12.7, and 20.32~cm, respectively.}
\label{fig:F_max_scaling}
\end{figure}

Figure~\ref{fig:F_max_scaling}(a) shows typical experimental trajectories. We plot $v$ and $-Ma$ with respect to time after impact, and then record $F_{\rm max} = - Ma_{\rm max}+Mg$, where $g=9.81$~m/s$^2$, and $t_{\rm max}$ as a function of $v_0$, as plotted in Fig.~\ref{fig:F_max_scaling}(b,c). These reveal power law scaling for $F_{\rm max}$ and $t_{\rm max}$ versus $v_0$: $F_{\max}\propto v_0^{4/3}$ and $t_{\max}\propto v_0^{-2/3}$. At small $v_0$, $F_{\rm max}$ appears to plateau as expected since $F_{\rm max}\approx Mg$ for very slow impacts where the granular force will increase until it approximately balances the gravitational force. These measurements for soft and medium particles are unambiguous (photoelastic and tracking data agree well). Fast force dynamics for the hard particles, with $v_0/v_b<10^{-2}$, cause $F_{\rm max}$ to be underestimated from only tracking the intruder~\cite{clark_prl12}, as seen in Fig.~\ref{fig:F_max_scaling}(a). The force from tracking the intruder [the thicker, black line in Fig.~\ref{fig:F_max_scaling}(a)] has a clear peak slightly above 10~N during $0<t<0.01$~seconds, while the photoelastic signal [the thinner, black line in Fig.~\ref{fig:F_max_scaling}(a)] shows a burst of peak forces: one above 50~N and four more around 30~N during this time. This is typical of all force measurements during impacts into hard particles. Thus, the data for hard particles in Fig.~\ref{fig:F_max_scaling}(b) is measured from video tracking and then multiplied by a constant correction factor (roughly 4.5) to account for this difference. The largest peak in the photoelastic signal is not always the first one, so we measure $t_{\rm max}$ from the video tracking data [e.g., $t_{\rm max} \approx 0.005$ in Fig.~\ref{fig:F_max_scaling}(a)], which represents the mean time associated with the burst of large forces observed in the photoelastic signal. Similar results are found by measuring $F_{\rm max}$ and $t_{\rm max}$ directly from the calibrated photoelastic data, but with significantly more scatter. The peak forces measured in this way for hard particles is similar to the data for soft and medium particles, as shown in Fig.~\ref{fig:F_max_scaling}(b,c), albeit with slightly different phenomenology: a burst of peak forces rather than a clear buildup and relaxation (this is also evident from Supplemental Movies in Ref.~\cite{clark_prl15}). This suggests a possible change in behavior for $v_0/v_b \ll 10^{-2}$. Previous studies using even stiffer grains~\cite{goldman_pre08,aguilar2016} have sometimes found $F_{\rm max}\propto v_0^2$ for similar impact speeds $v_0\sim 1$ to 5~m/s into, e.g., glass beads where $E^\ast \sim 50$ to 100 GPa, $\rho_g \sim 2000$~kg/m$^3$, and thus $v_0/v_b<10^{-3}$.

\begin{figure}
\raggedright (a) \hspace{40mm} (b) \\ \centering
\includegraphics[trim=0mm 0mm 0mm 0mm, clip, width=0.49\columnwidth]{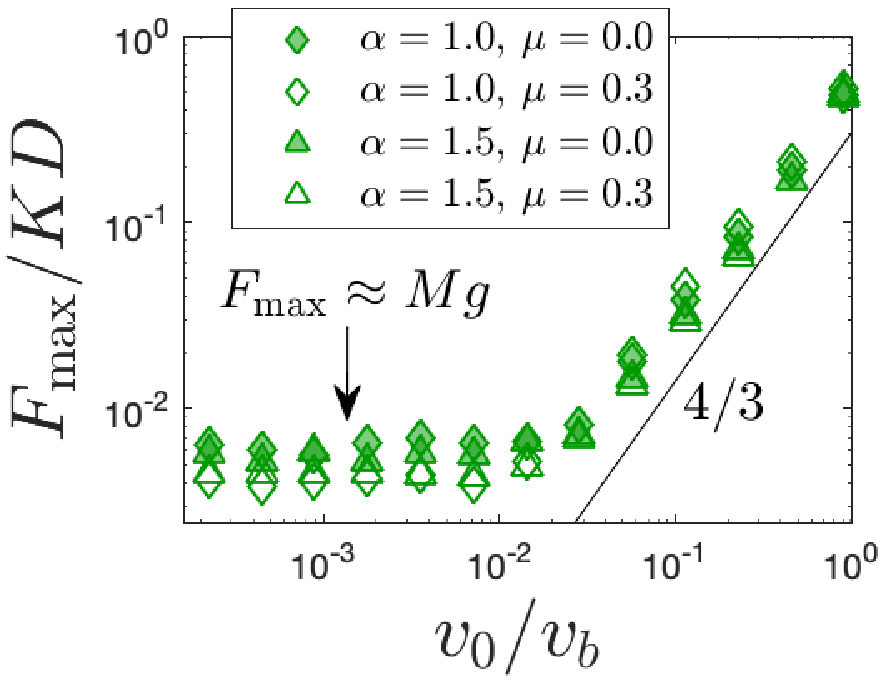}
\includegraphics[trim=0mm 0mm 0mm 0mm, clip, width=0.49\columnwidth]{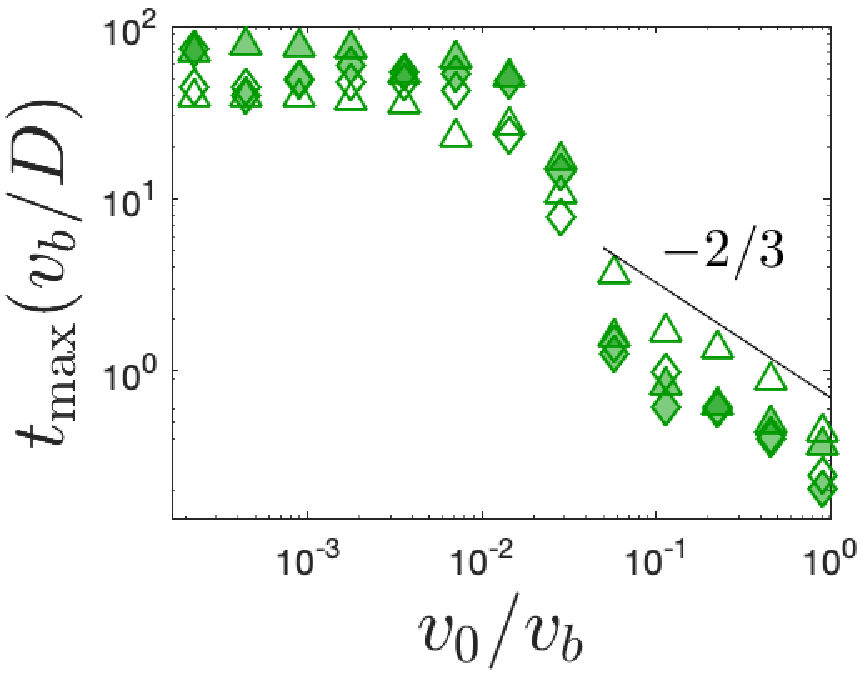}
\raggedright (c) \hspace{40mm} (d) \\ \centering
\includegraphics[trim=0mm 0mm 0mm 0mm, clip, width=0.49\columnwidth]{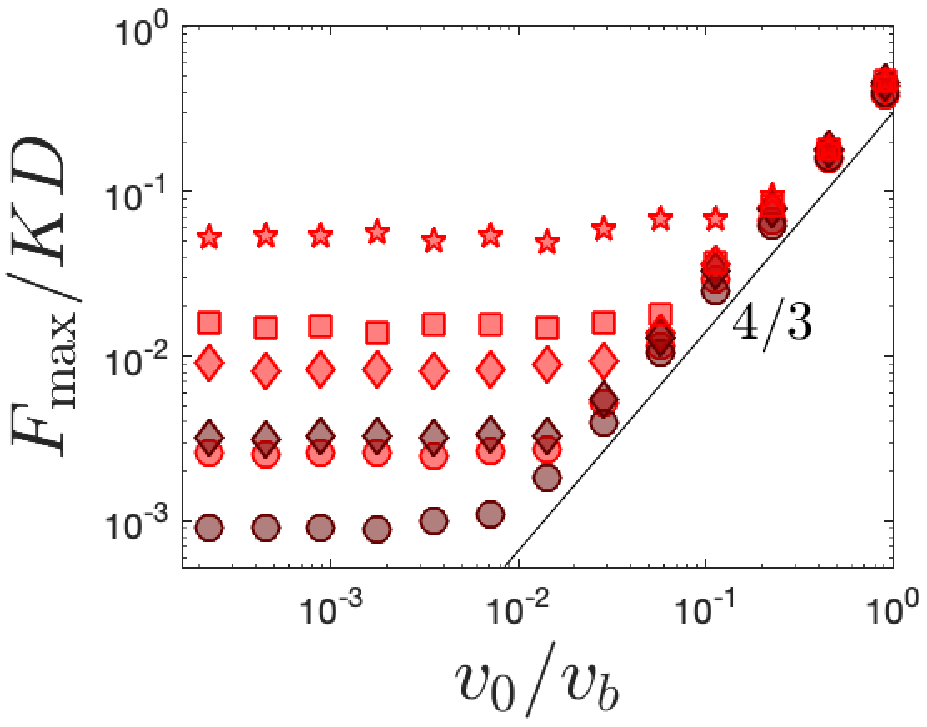}
\includegraphics[trim=0mm 0mm 0mm 0mm, clip, width=0.49\columnwidth]{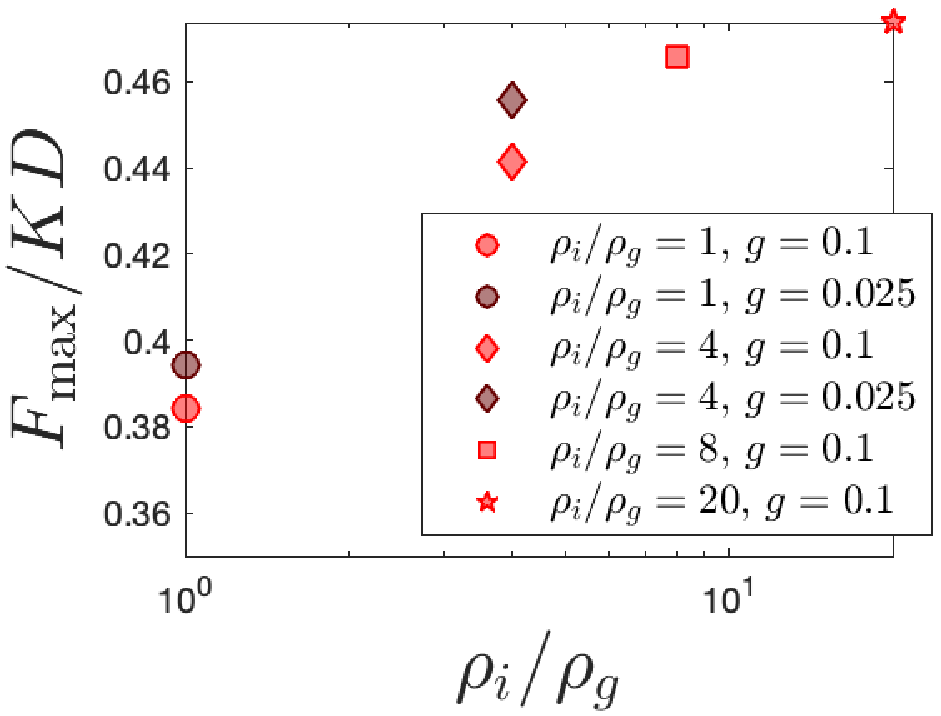}
\caption{(a) $F_{\rm max}/KD$ and (b) $t_{\rm max}(v_b/D)$ are plotted as a function of $v_0/v_b$ for 2D simulations with $D=7$, $d=1$, $\sigma_g = \rho_g w =1$, $K=785$, $\sigma_i = \rho_i w = 4$, and $g=0.1$ with varying grain-grain friction coefficient $\mu$ and force law exponent $\alpha$, showing that $F_{\rm max}\propto v_0^{4/3}$ and $t_{\rm max}\propto v_0^{-2/3}$ are independent of $\mu$ and $\alpha$. The legend in (a) also applies to (b). (c) $F_{\rm max}/KD$ is plotted as a function of $v_0/v_b$ for simulations with $D=14$, $d=1$, $\sigma_g =\rho_g w=  1$, $K=785$, $\mu = 0$, $\alpha = 1$. The power-law scaling is independent of varied gravitational constant $g$ and intruder mass density $\rho_i/\rho_g$; the plateau value at low $v_0$ is approximately equal to $Mg$, as expected. (d) $F_{\rm max}/KD$ is plotted as a function of $\rho_i/\rho_g$ for the largest value of $v_0$ shown in (c), $v_0 \approx v_b$, showing that intruder weight has little effect on the peak force, especially when $\rho_i>4\rho_g$. The legend in (d) applies to (c) as well.}
\label{fig:F_max_params}
\end{figure}

To better understand the origins of the power law behavior in Fig.~\ref{fig:F_max_scaling}(b,c), we perform discrete-element simulations~\cite{cundall79,clark_pre16} using C++ in 2D and LAMMPS~\cite{plimpton1995fast} (http://lammps.sandia.gov) in 3D; further details are given in Supplemental Material. We prepare a static, gravitationally loaded bed of 10,000 grains in 2D and 100,000 grains in 3D, which we verify are large enough that system boundaries (lateral or bottom) do not affect our results in any way. Previous work has sometimes found that the initial packing fraction plays a role in the impact response~\cite{umbanhowar_pre10,jerome2016} due to dilation (for densely prepared systems) or compaction (for loosely prepared systems) during shear. This effect is suppressed for frictionless grains~\cite{Peyneau2008}. Our results are not sensitive to the initial packing fraction of the bed or, as shown below, even to the existence of friction at all. Thus, we conclude that shear-induced dilation and compaction do not affect the scaling laws we show. After preparing the bed, we then put a circular (2D) or spherical (3D) intruder just above the bed with downward velocity $v_0$, after which it is free to accelerate due to forces from grains or gravity. We observe trajectories that are similar to those shown in Figs.~\ref{fig:soft-frames} and \ref{fig:F_max_scaling}(a) as well as throughout the literature~\cite{picaciamarra_prl04,tiwari2014drag}. We again find $F_{\rm max} \propto v_0^{4/3}$ and $t_{\rm max}\propto v_0^{-2/3}$, as shown in Fig.~\ref{fig:F_max_params}. At slow speeds, the power-law scaling is cut off by $F_{\rm max} \approx Mg$ since gravity accelerates the intruder and generates forces comparable to its weight. This occurs at longer $t_{\rm max}$, nearly independent of $v_0$.

\begin{figure}
\raggedright (a) \hspace{40mm} (b) \\ \centering
\includegraphics[trim=0mm 0mm 5mm 0mm, clip, width=0.49\columnwidth]{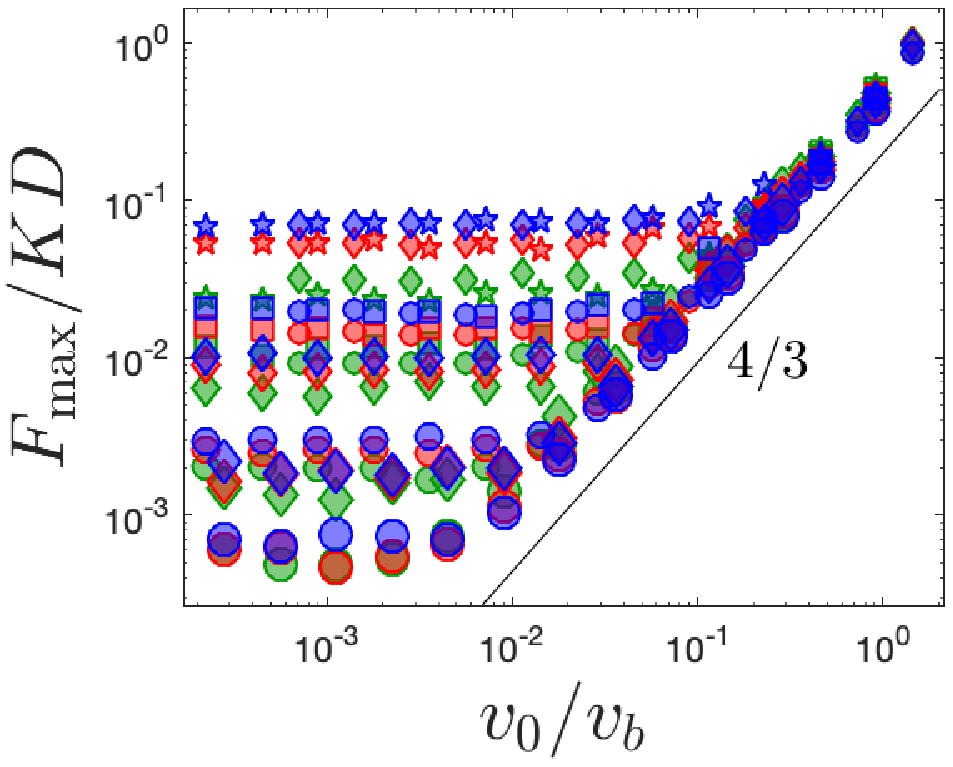}
\includegraphics[trim=0mm 0mm 5mm 0mm, clip, width=0.49\columnwidth]{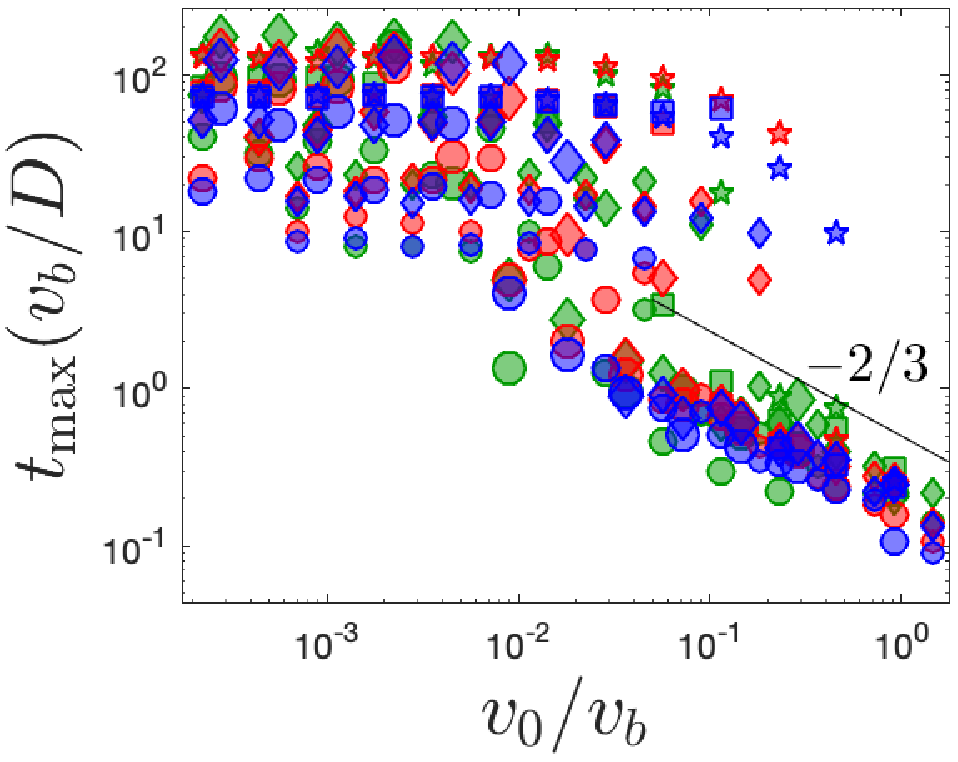}
\raggedright (c) \hspace{40mm} (d) \\ \centering
\includegraphics[trim=0mm 0mm 5mm 0mm, clip, width=0.49\columnwidth]{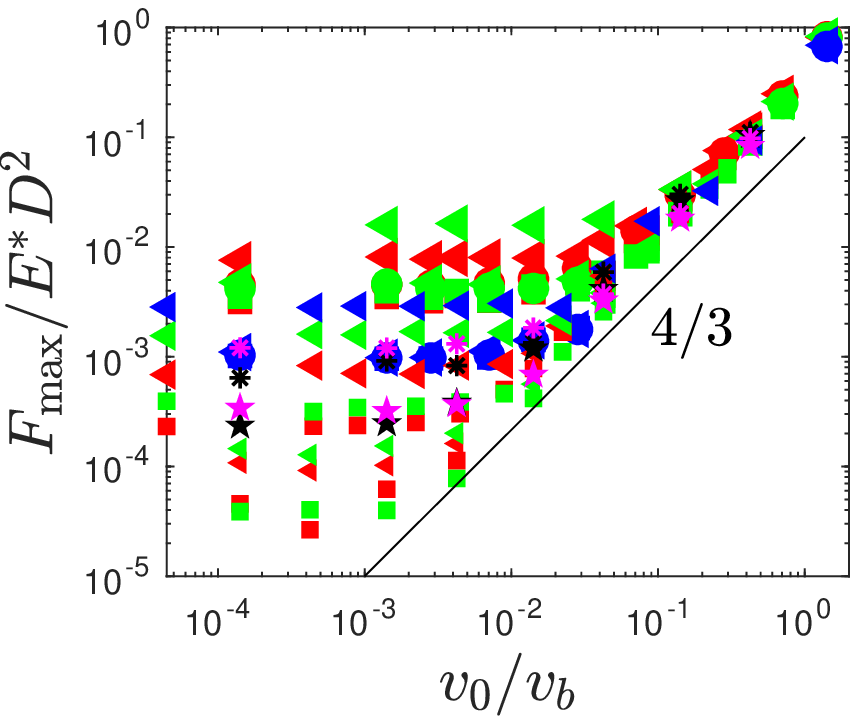}
\includegraphics[trim=0mm 0mm 5mm 0mm, clip, width=0.49\columnwidth]{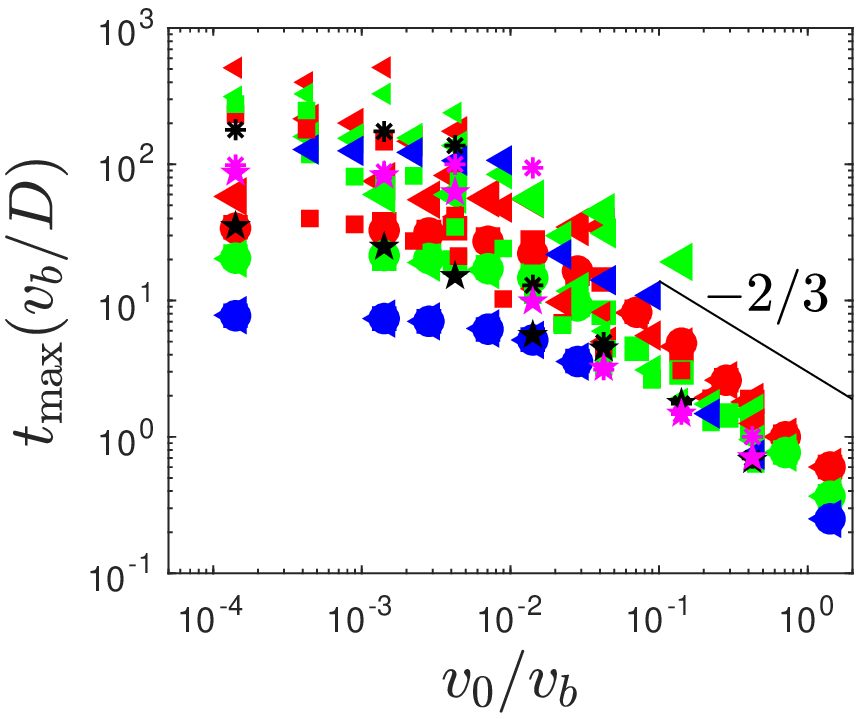}
\raggedright (e) \hspace{40mm} (f) \\ \centering
\includegraphics[trim=0mm 0mm 5mm 0mm, clip, width=0.49\columnwidth]{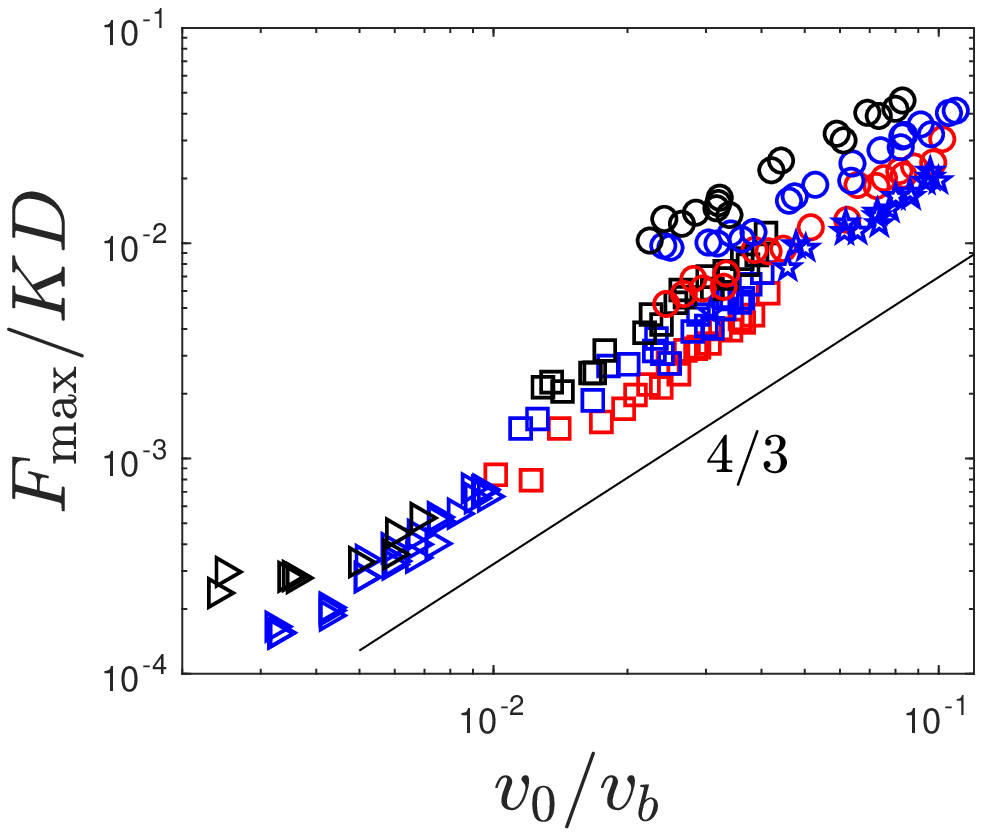}
\includegraphics[trim=0mm 0mm 5mm 0mm, clip, width=0.49\columnwidth]{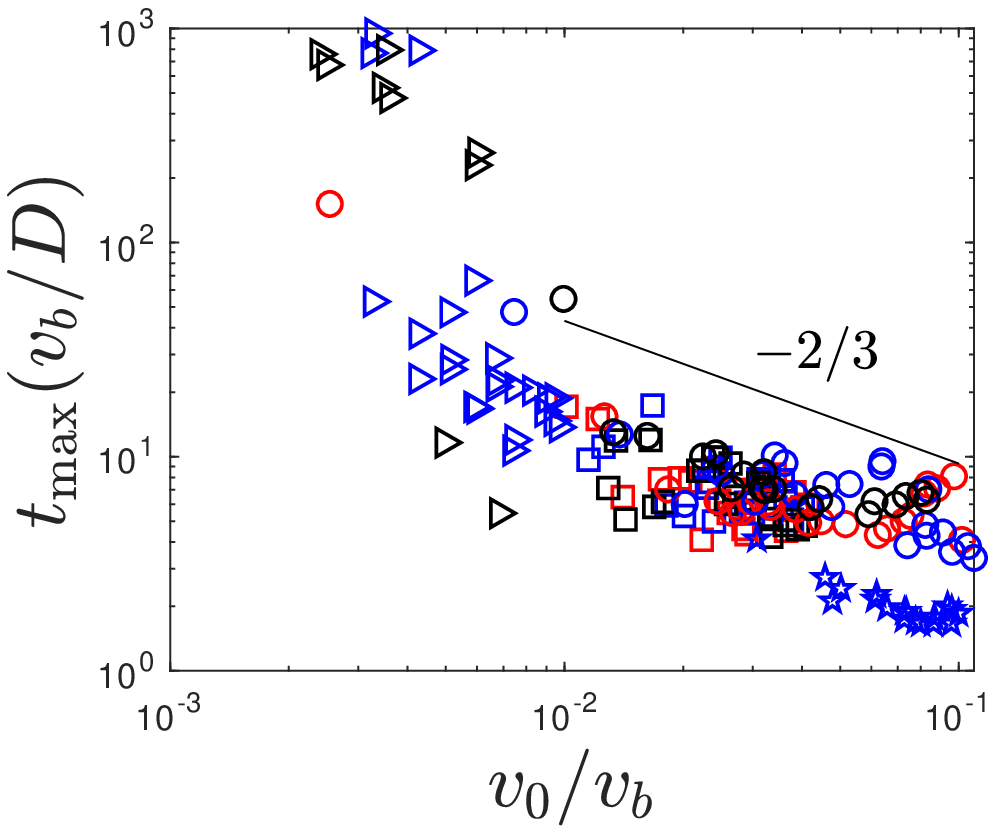}

\caption{Scaled plots of $F_{\rm max}$ and $t_{\rm max}$ versus $v_0$ for (a,b) simulations of 2D, frictionless grains with $\alpha=1$, (c,d) simulations of 3D frictional grains with $\alpha = 1.5$, and (e,f) 2D, frictional grains from experiment, with $\alpha \approx 1.4$. The symbols in (e,f) are the same as Fig.~\ref{fig:F_max_scaling}(c,d); For simulations shown in (a-d), the full symbol list is given in Supplemental Material. Plateau values for small $v_0/v_b$ are set by $F_{\rm max} \approx Mg$.}
\label{fig:F_max_scaled_all}
\end{figure}

For 2D simulations of circular intruders impacting beds of frictional, circular grains, there are nine system parameters: the intruder diameter $D$, mass $M$, and speed $v_0$; the grain diameter $d$, mass $m$, stiffness $K = E^\ast w$, friction coefficient $\mu$, and force law exponent $\alpha$; and the gravitational constant $g$. Masses per area for grains and intruder are $\sigma_i = \rho_i w = 4M/\pi D^2$ and $\sigma_g = \rho_g w = 4m/\pi d^2$, where $\rho_i$ and $\rho_g$ are the masses per volume and $w$ is the thickness of the particles ($w$ only has meaning in the experiments). Figure~\ref{fig:F_max_params}(a,b) show that the power law scaling is nearly independent of both $\alpha$ and $\mu$. The lack of dependence on $\mu$ suggests that dilation or compaction due to shear, which only occurs for frictional grains, does not affect our results (possibly because the material does not have time to develop any shear-induced dilation or compaction). Figure~\ref{fig:F_max_params}(c,d) show that the intruder weight $Mg$ sets the value of the plateau, as expected, but does not affect the forces due to the power-law scaling, particularly once $\sigma_i / \sigma_g > 4$. We also find our results do not explicitly depend on $d$; this is implicitly shown in Fig.~\ref{fig:F_max_scaled_all}, which includes values of $d$ that vary by an order of magnitude in 3D (however, we only study cases with $D\geq 5d$). The remaining quantities $F_{\rm max}$, $t_{\rm max}$, $v_0$, $D$, $K$, and $\sigma_g$ form three dimensionless groups, $F_{\rm max}/KD$, $t_{\rm max} v_b / D$, and $v_0 / v_b$, where $v_b = \sqrt{K/\sigma_g} = \sqrt{E^\ast / \rho_g}$. Figure~\ref{fig:F_max_scaled_all}(a,b) explicitly shows that all our data for 2D frictionless Hookean simulations collapse when $F_{\rm max}/KD$ and $t_{\rm max} v_b / D$ are plotted as a function of $v_0 / v_b$. We note that the collapse in Figure~\ref{fig:F_max_scaled_all}(a,b) includes data spanning two orders of magnitude in $K$, which was not explicitly shown in Fig.~\ref{fig:F_max_params} but is shown in Supplemental Material. 

In 3D, grain-grain compression is governed by $f=E^\ast d^2\left(\frac{\delta}{d}\right)^\alpha$ (along with frictional and dissipative intergrain interactions; see Supplemental Material). We again find $F_{\rm max} \propto v_0^{4/3}$ and $t_{\rm max}\propto v_0^{-2/3}$, with a plateau at slow speeds set by $F \approx Mg$. The power-law behavior is again nearly independent of $\mu$, $\alpha$, $g$, $d$, and the mass per volume $\rho_i$ of the spherical intruder (provided $\rho_i>\rho_g$). The remaining parameters $F_{\rm max}$, $t_{\rm max}$, $v_0$, $E^\ast$, $D$, and $\rho_g$ form three dimensionless groups: $F_{\rm max}/E^\ast D^2$, $t_{\rm max} v_b / D$, and $v_0/v_b$, where $v_b = \sqrt{E^\ast/\rho_g}$. Figure~\ref{fig:F_max_scaled_all}(c,d) shows collapsed data in 3D frictional Hertzian simulations, which are nearly identical to the 2D frictionless Hookean results shown in Figure~\ref{fig:F_max_scaled_all}(a,b). Experimental results, plotted in Fig.~\ref{fig:F_max_scaled_all}(e,f), are more scattered than the simulations; this is partly expected since force is not directly measured but inferred by tracking the intruder and, in the case of hard particles, using a correction from a calibrated photoelastic signal. Experimental impacts of large intruders into soft particles appear to deviate from the scaling, which could be due to the extreme particle deformation and collective stiffening during these impacts~\cite{clark_prl15}.

Thus, we find universal scaling laws for the peak forces and associated time scales during the early stage of high-speed impact: $F_{\rm max} \propto KD (v_0 / v_b)^{4/3}$ in 2D, $F_{\rm max} \propto E^\ast D^2(v_0 / v_b)^{4/3}$ in 3D, and $t_{\rm max} \propto (D/v_b)(v_0 / v_b)^{-2/3}$ in both 2D and 3D. As we now show, these scaling laws are inconsistent with Poncelet, shock, and added mass models, suggesting that a new theory must be formulated. The scaling $F_{\max}\propto v_0^{4/3}$ is plainly inconsistent with Poncelet drag, where $F \propto v^{2}$. Granular shock theory~\cite{nesterenko95,daraio_pre05,Job05,daraio_pre06,gomez_prl12,vandenWildenberg_prl13}, which captures the speed of propagating forces in these experiments~\cite{clark_prl15}, is fundamentally based on the force exponent $\alpha$ and states that large stresses (compared to the prestress) propagate at a shock speed $v_f$, where $(v_f/v_b)\propto (v_0/v_b)^{ \frac{\alpha - 1}{\alpha + 1} } \propto \left(P_{\rm max}\right)^{ \frac{\alpha - 1}{2 \alpha} }$. By rearranging, $P_{\rm max} \propto v_0^{ \frac{2 \alpha}{\alpha + 1} }$, which yields $P_{\rm max} \propto v_0^{6/5}$ for Hertzian grains (where $\alpha = 1.5$) and $P_{\rm max} \propto v_0^{7/6}$ for the disks we use (with $\alpha \approx 1.4$), which are also inconsistent with $F_{\rm max}\propto v_0^{4/3}$. Even a modified shock theory that somehow predicted $F_{\rm max}\propto v_0^{4/3}$ must be fundamentally based on $\alpha$. Thus, the lack of dependence on $\alpha$ shown in Fig.~\ref{fig:F_max_params}(a,b) confirms that granular shock theory cannot explain our results. 

A third possible explanation is added-mass effects~\cite{Truscott2014,waitukaitis2012impact,aguilar2016,Mukhopadyay2018}. Added-mass models assume that the intruder, with mass $M$, decelerates primarily due to rigid connection to a growing mass of the material, with mass $m_a(t)$. Assuming external forces are known, a mathematical form for $m_a(t)$ will then fully determine the trajectory, including $F_{\rm max}$. This scenario is consistent with the images shown in Fig.~\ref{fig:soft-frames}, since there is a growing cluster of grains that is connected to the intruder. For comparison, we numerically solve added-mass dynamics (see Supplemental Material for details), assuming the added mass is a half circle (in 2D) or sphere (in 3D) with a radius $R$ that grows at the force propagation speed $v_f$, i.e., $R = v_f t$, as suggested by Fig.~\ref{fig:soft-frames}. This gives $m_a(t) = \pi (v_f t)^2/2$ in 2D and $m_a(t) = 2\pi (v_f t)^3/3$ in 3D, with $v_f$ as a constant (true for a Hookean force law). Solving this model yields $F_{\rm max} = A v_0$ in 2D and 3D, not $F_{\rm max}\propto v_0^{4/3}$; $A\propto M^{1/2}$ in 2D and $A\propto M^{2/3}$ in 3D; and $t_{\rm max}$ independent of $v_0$, not $t_{\rm max} \propto (v_0)^{-2/3}$. It is possible that $F_{\rm max} \propto v_0^{4/3}$ could be obtained for some choice $m_a(t)$, but $A$ always increases with $M$ for the forms of $m_a(t)$ that we try, as well as in the theoretical analysis in Ref.~\cite{Mukhopadyay2018} of the added-mass model from Ref.~\cite{waitukaitis2012impact}. This is inconsistent with the lack of dependence on $\rho_i$, shown in Fig.~\ref{fig:F_max_params}(c,d). 

We note that the soft repulsive disks and spheres used in the simulations are commonly used to model other soft, particulate media (like foams or emulsions), suggesting that this description will likely apply to a much broader group of materials. We also note that the maximum grain compression $\delta_{\rm max}$ can be easily estimated from the scaling laws we show, which could be used to predict grain fracture or crushing. In 3D, with $P_{\rm max} = F_{\rm max}/D^2$, the maximum force felt by a grain is $f_{\rm max} \sim P_{\rm max}d^2 \sim d^2 E^\ast (v_0 / v_b)^{4/3}$. For the Hookean force law, $f_{\rm max} = E^\ast d \delta_{\rm max}$, so $\delta_{\rm max}/d \sim (v_0/v_b)^{4/3}$. This is why we do not show any impacts with $v_0>v_b$: grain-grain overlaps in simulations became similar to the size of a grain. Physical grains would certainly be crushed or fractured in this regime, and other physics would likely become dominant.

\begin{acknowledgments}
We acknowledge funding from the Office of Naval Research under Grant No. N0001419WX01519. These experiments were performed in the late R. P. Behringer's lab during dissertation work by AHC. We also thank Jeffrey Haferman and Bruce Chiarelli for help with high-performance computing at NPS.
\end{acknowledgments}


\end{document}